\documentstyle[12pt,epsf]{article}

\def\SO{\mathop{\rm SO}}
\def\SU{\mathop{\rm SU}}
\def\U{\mathop{\rm U}}

\begin{document}

\begin{titlepage}
    \begin{normalsize}
     \begin{flushright}
                 UT-03-27\\
                 {\tt hep-th/0309024}\\
                 September 2003
     \end{flushright}
    \end{normalsize}
    \begin{Large}
       \vspace{0.6cm}
       \begin{center}
       \bf Decay of type 0 NS5-branes to nothing
       \end{center}
    \end{Large}
  \vspace{15mm}

\begin{center}
{\large
           Yosuke Imamura
           \footnote{E-mail :
              imamura@hep-th.phys.s.u-tokyo.ac.jp}

      \vspace{16mm}

              {\it Department of Physics,}\\[0.7ex]
              {\it University of Tokyo,}\\[0.7ex]
              {\it Hongo, Tokyo 113-0033, Japan}
}
      \vspace{1.5cm}
\end{center}
\begin{abstract}
The perturbative vacuum of type 0 string theory
is unstable due to the existence of
the closed string tachyon.
This instability can be removed by ${\bf S}^1$ compactification
with twisted boundary condition for the tachyon field.
We show that even in this situation unwrapped
NS5-branes are unstable and
decay to bubbles of nothing smoothly without tunneling
any potential barrier.
We discuss a relation between the closed string tachyon condensation
and the instability of NS5-branes.
\end{abstract}

\end{titlepage}

\section{Introduction}
Tachyon condensation is one of most interesting and important
subject in the recent development in string theory.
After the connection between the open string tachyon field
and unstable D-branes was conjectured by Sen\cite{Sen1,Sen2,Sen3,Sen4},
our understanding of dynamics of D-branes has been greatly
deepened.
Having looked at this success, many people anticipate that the
closed string tachyon field also has clues to non-perturbative dynamics
of string theory.
Although the role of closed string tachyon condensation in string theory
is not as clear as that of open string,
several ideas have been proposed.
In \cite{Melvin,fluxbrane}, relation between instability of fluxbranes
and closed string tachyon condensation is discussed.
In \cite{panic}, it is suggested that
condensation of tachyonic twisted modes
localized at fixed points of non-supersymmetric orbifolds
like ${\bf C}/{\bf Z}_n$ and ${\bf C}^2/{\bf Z}_n$
give rise to deformations of geometries of the background
spacetimes.
Configurations bringing with localized tachyons are also discussed in
many works\cite{tseytlin,dabholkar,vafa,HKMM,michiyi,twistedcircle,namsin,deAlwisFlournoy,RabadanSimon,uranga,dacunhamartinec}.

In this paper, we study localized tachyonic modes
on a type 0 NS5-brane.
Type 0 string theory includes the closed string tachyon field in the bulk.
In order to remove the instability associated with it,
we compactify one spacelike direction of the background spacetime
on ${\bf S}^1$ with a twisted boundary condition for
left-moving worldsheet fermions.
Because the bulk tachyon field changes its sign under this twist,
it becomes massive when the compactification
radius is sufficiently small.
We will show in this paper that even in such a situation
there are localized tachyonic modes on an unwrapped NS5-brane
and it decays into a bubble of nothing\cite{kkvacuum}.

One way to obtain a spectrum of fields localized on an NS5-brane
is to analyze fluctuations around a corresponding classical solution
of the Einstein equation.
In \cite{CrapsRoose},
a type 0 NS5-brane is mapped to a type 0 Kaluza-Klein monopole
by using type 0/type 0 T-duality\cite{BG},
and a non-chiral spectrum on an NS5-brane is obtained
as zero-modes of massless fields in type 0 string theory.

In Section 2 and 3, we adopt a similar way
to show the existence of a tachyonic mode on an NS5-brane.
Because we assume twisted boundary condition on ${\bf S}^1$ for the bulk tachyon,
T-duality maps an unwrapped type 0 NS5-brane into
type II string theory with a spacetime geometry similar to a two-centered
Kaluza-Klein monopole\cite{ima1}.
We describe it in detail in the next section,
and show in Section 3 that it decays with creating a bubble of nothing at the center of the manifold.

In Section 4, we analyze fluctuations of the closed string
tachyon field on an NS5-brane classical solution and show that
there exist localized tachyonic modes, however small
the compactification radius.
We will find that its tachyonic mass is of the same order
with what obtained in Section 3 and
we conjecture that the closed string tachyon condensation
on an NS5-brane is responsible for the instability of the geometry.

The last section is devoted for discussions.
We mention there about similarity and difference between our geometry
and monopole-anti monopole pair creation
studied in other works.

\section{Construction of a composite manifold}
In this section we construct a four-dimensional manifold
such that ${\bf R}^6\times{\cal M}$ is the T-dual of a unwrapped type 0
NS5-brane by combining two hyper-K\"ahler manifolds.
It is known that there are two smooth four-dimensional
hyper-K\"ahler manifolds with locally $\SU(2)\times\U(1)$ isometry\cite{EGH,AH}.
The word `locally' means that at this point we do not distinguish
$\SO(3)$ and $\SU(2)$.
One of two manifolds is Taub-NUT manifold (${\cal N}_{\rm TN}$)
and the other is Eguchi-Hanson manifold (${\cal N}_{\rm EH}$).
By looking at the global structure of these manifolds,
we find that isometry of ${\cal N}_{\rm EH}$ is $\SO(3)\times\U(1)$
while that of ${\cal N}_{\rm TN}$ is $\SU(2)\times\U(1)$.
Correspondingly, isometry orbits for generic points are
topologically ${\bf S}^3/{\bf Z}_2$ for ${\cal N}_{\rm EH}$ and
${\bf S}^3$ for ${\cal N}_{\rm TN}$.
These four-dimensional manifolds are parameterized by
three angular coordinate on ${\bf S}^3$ or ${\bf S}^3/{\bf Z}_2$
and radius $r$.
We define the coordinate $r$ as a geodesic distance from
the center of the manifolds.
(The ``center'' means NUT for ${\cal N}_{\rm TN}$ or bolt for ${\cal N}_{\rm EH}$.)

In what follows, we construct a ``composite manifold'' $\cal M$ by pasting
the large $r$ part of ${\cal N}_{\rm TN}$ and
the small $r$ part of ${\cal N}_{\rm EH}$.
In order for two manifolds to join smoothly, the topology of their sections
must be the same and the metrics of two manifolds
should be appropriately deformed so as to interpolate inside and outside
smoothly.
To obtain the same topology of the sections
of two manifolds, we will use ${\cal N}_{\rm TN}/{\bf Z}_2$ rather than
${\cal N}_{\rm TN}$ itself.
(${\bf Z}_2$ is the center of the $\SU(2)$ factor in the isometry group.)
Then, both the sections are ${\bf S}^3/{\bf Z}_2$,
and
these can be represented as an ${\bf S}^1$ fibration over ${\bf S}^2$
(Hopf fibration) with first Chern class $2$.
Although the orbifold ${\cal N}_{\rm TN}/{\bf Z}_2$ is singular at its
center, it does not matter because we will only use the large $r$ part
of the manifold.

Each metric of two manifolds ${\cal N}_{\rm TN}/{\bf Z}_2$
and ${\cal N}_{\rm EH}$ includes one scale parameter.
We denote them by $a_{\rm TN}$ and $a_{\rm EH}$, respectively.
For ${\cal N}_{\rm TN}/{\bf Z}_2$, $a_{\rm TN}$ represents the radius of
${\bf S}^1$ fiber at $r=\infty$.
On the other hand, $a_{\rm EH}$ represents the size of
${\bf S}^2$ at the bolt (the place where ${\bf S}^1$ fiber shrinks to a point)
of ${\cal N}_{\rm EH}$.
Let us consider a situation where $a_{\rm TN}$ is much larger
than $a_{\rm EH}$.
In this situation, there is an intermediate scale $a_0$
satisfying $a_{\rm EH}\ll a_0\ll a_{\rm TN}$.
If we look at the intermediate region $r\sim a_0$,
both these manifolds are approximately ${\bf R}^4/{\bf Z}_2$.
Therefore, we can construct a composite manifold ${\cal M}$
by combining $r\leq a_0$ part of ${\cal N}_{\rm EH}$
and $a_0\leq r$ part of ${\cal N}_{\rm TN}/{\bf Z}_2$.
Of cause we need to deform two manifolds to obtain
a smooth manifold. But the deformation becomes smaller as the ratio
$a_{\rm TN}/a_{\rm EH}$ becomes larger.

If we start from an orbifold ${\bf R}^4/{\bf Z}_2$ around the intermediate
region and extend it to interior ${\cal N}_{\rm EH}$ or
to exterior ${\cal N}_{\rm TN}/{\bf Z}_2$,
we can choose one of two ``orientations''.
One way to see this is to look at the relation between isometry
of the intermediate region and that of the internal or external region.
The isometry of ${\bf R}^4/{\bf Z}_2$ is $\SO(3)_R\times\SO(3)_L$.
By the extension to the internal or external region,
this symmetry is broken to $\SO(3)\times\U(1)$.
There are two choices of an unbroken $\SO(3)$ from $\SO(3)_L\times\SO(3)_R$
and these correspond to two orientations of extension.
We refer to an extension as ``left-handed'' if it preserves $\SO(3)_L$,
and the opposite extension as ``right-handed''.
Because we have two choices for both internal and external extensions,
we obtain four different composite manifolds.
However, we are interested only in the relative orientation
and there are two essentially different composite manifolds.

Each constituent manifold ${\cal N}_{\rm TN}/{\bf Z}_2$ or ${\cal N}_{\rm EH}$
possesses three complex structures $I^{(a)}$ ($a=1,2,3$)
satisfying
$I^{(a)}I^{(b)}=-\delta_{ab}+\epsilon_{abc}I^{(c)}$.
We call a set of these three complex structures hyper-K\"ahler structure.
It is important to know if it is possible to preserve the
hyper-K\"ahler structure on the composite manifold
after an appropriate deformation of the metric.
To obtain some information about this problem,
let us think of the relation between isometries and hyper-K\"ahler structures.
An important property of the hyper-K\"ahler structure on
${\cal N}_{\rm TN}/{\bf Z}_2$ and that on ${\cal N}_{\rm EH}$
is that the former belongs to $\bf 3$ of $\SO(3)$ isometry while
the latter is an $\SO(3)$ singlet.
From this fact, we can make a guess as follows.
If a composite manifold consists of
two manifolds with the same orientation (left-handed$+$left-handed or right-handed$+$right-handed),
the same $\SO(3)$ is preserved by the internal and external regions.
This means that two parts of the manifold respect different hyper-K\"ahler structures.
Therefore, the entire manifold would not possess any hyper-K\"ahler structure.
We call this manifold $\SO(3)$-symmetric composite manifold
and denote it as ${\cal M}_{\SO(3)}$.
On the other hand, if the orientations of the internal and external parts are
opposite
(left-handed$+$right-handed or right-handed$+$left-handed),
although the two parts respect the different $\SO(3)$
isometries, these respect the same hyper-K\"ahler structure.
This strongly suggests the existence of an appropriate deformation of
metric preserving a hyper-K\"ahler structure.
Indeed, we can identify this manifold with a two-centered Taub-NUT manifold and
the parameter $a_{\rm EH}$ determines the distance between the two centers.
We call this manifold hyper-K\"ahler composite manifold
and denote it as ${\cal M}_{\rm HK}$.

We can represent ${\cal M}_{\SO(3)}$ and ${\cal M}_{\rm HK}$ as ${\bf S}^1$ fibrations on
asymptotically flat three-dimensional base manifolds.
In the hyper-K\"ahler case,
as we mentioned above,
the manifold ${\cal M}_{\rm HK}$ is a two-centered Taub-NUT, and the base manifold is ${\bf R}^3$.
For $\SO(3)$-symmetric manifold ${\cal M}_{\SO(3)}$, the base manifold is parameterized by
two angular coordinates on ${\bf S}^2$ (base manifold of Hopf fibration of
${\bf S}^3/{\bf Z}_2$) and the radius of the ${\bf S}^2$.
Because the radius is bounded below by $a_{\rm EH}$,
the base manifold is ${\bf R}^3$ with solid ${\bf S}^2$ with
radius $a_{\rm EH}$ omitted.
This is a ``bubble of nothing''\cite{kkvacuum}.

Let us consider compactifications of type II string theory on these
composite manifolds.
As we mentioned above, the composite manifolds ${\cal M}_{\SO(3)}$ and ${\cal M}_{\rm HK}$
have the same asymptotic form
at large $r$.
They are ${\bf S}^1$-fibrations over flat ${\bf R}^3$.
Because the ${\bf S}^1$-cycles are shrinkable
in both manifolds,
the boundary conditions around the ${\bf S}^1$ for fermions are
uniquely determined.
We can easily show that it is periodic for ${\cal M}_{\rm HK}$ and
anti-periodic for ${\cal M}_{\SO(3)}$\cite{ima1}.
By the T-duality along the ${\bf S}^1$, these configurations are transformed to
${\bf S}^1$ compactified type II theory (${\cal M}_{\rm HK}$) and
${\bf S}^1$ compactified type 0 theory (${\cal M}_{\SO(3)}$)\cite{BG}.
The central parts of the manifolds are mapped to unwrapped NS5-branes in both cases.
A detailed consideration of the relation of
compactification radii, string winding numbers and Kaluza-Klein momenta
shows that the number of NS5-branes is two for ${\cal M}_{\rm HK}$
and one for ${\cal M}_{\SO(3)}$\cite{ima1}.
By taking advantage of this duality,
we analyze the stability of a type 0 NS5-brane in the next section.

\section{Gravitational instability}
In this section we analyze the stability of the $\SO(3)$-symmetric
composite manifold ${\cal M}_{\SO(3)}$, which is the T-dual to a type 0 NS5-brane.
We keep the parameter $a_{\rm TN}$ fixed as a boundary condition
because it represents the
radius of the ${\bf S}^1$ fiber at $r\rightarrow\infty$.
The other scale parameter $a_{\rm EH}$ is treated
as a deformation parameter of the manifold.
We assume here
that the background geometry is ${\bf R}^6\times{\cal M}_{\SO(3)}$,
the dilaton is constant, and all the other fields vanish.
By taking a certain ansatz for the metric of the composite manifold,
and computing Einstein-Hilbert Lagrangian,
we can define an effective potential
$V(a_{\rm EH})$ and the configuration turns out to be unstable.
Therefore, the parameter $a_{\rm EH}$ plays a role of a tachyon field
localized near the center of the manifold.
Because $a_{\rm EH}$ also represents the radius of a bubble,
we conclude that a type 0 NS5-brane decays into a bubble of nothing.

\begin{figure}[htb]
\centerline{\epsfbox{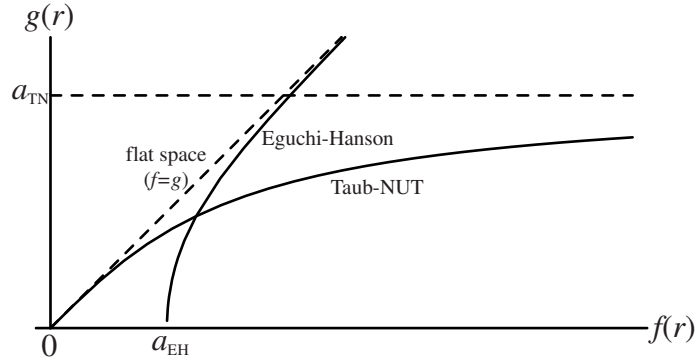}}
\caption{Parametric plots of functions $f(r)$ and $g(r)$ for
Eguchi-Hanson and Taub-NUT manifolds.}
\label{gplot.eps}
\end{figure}

For each value of $a_{\rm EH}$, we have a pair of ${\cal N}_{\rm EH}$
and ${\cal N}_{\rm TN}/{\bf Z}_2$.
Thanks to the $\SO(3)\times\U(1)$ isometry,
the metrics of both manifolds are written in the following form.
\begin{equation}
ds^2=dr^2+f^2(r)(\sigma_x^2+\sigma_y^2)+g^2(r)\sigma_z^2,
\label{fgmetric}
\end{equation}
where $\sigma_x$, $\sigma_y$, and $\sigma_z$ are the Maurer-Cartan one-forms
on $\SO(3)$ group manifold defined by
\begin{eqnarray}
\sigma_x&=&\cos\psi d\theta-\sin\psi\sin\theta d\phi,\nonumber\\
\sigma_y&=&\sin\psi d\theta+\cos\psi\sin\theta d\phi,\nonumber\\
\sigma_z&=&d\psi-\cos\theta d\phi.
\end{eqnarray}
The ranges of angular coordinates are
\begin{equation}
0\leq\theta\leq\pi,\quad
0\leq\phi\leq2\pi,\quad
0\leq\psi\leq2\pi.
\end{equation}
In the case of $\SU(2)$ group manifold, the period of $\psi$ is $4\pi$.
Now this is halved due to the ${\bf Z}_2$ orbifolding.
One way to determine the functions $f(r)$ and $g(r)$ for ${\cal N}_{\rm TN}/{\bf Z}_2$ and
${\cal N}_{\rm EH}$ is to solve the Einstein equation.
However, it is more convenient to use the fact that
the spin connection obtained from (\ref{fgmetric}) admits a hyper-K\"ahler structure.
Because hyper-K\"ahler structures of ${\cal N}_{\rm TN}/{\bf Z}_2$ and
${\cal N}_{\rm EH}$ are transformed under $\SO(3)$ isometry
in different ways,
first-order differential equations for two manifolds are different.
\begin{equation}
f_{\rm TN}'=1-\frac{g_{\rm TN}}{2f_{\rm TN}},\quad
g_{\rm TN}'=\frac{g_{\rm TN}^2}{2f_{\rm TN}^2},\quad
f_{\rm EH}'=\frac{g_{\rm EH}}{2f_{\rm EH}},\quad
g_{\rm EH}'=1-\frac{g_{\rm EH}^2}{2f_{\rm EH}^2}.
\label{killing}
\end{equation}
By solving these equations,
we obtain the following relations between $f$ and $g$ for each manifold.
(See also Figure \ref{gplot.eps}.)
\begin{equation}
g_{\rm EH}^2=f_{\rm EH}^2\left(1-\frac{a_{\rm EH}^4}{f_{\rm EH}^4}\right),\quad
g_{\rm TN}^2=a_{\rm TN}^2\frac{\sqrt{1+\frac{4f_{\rm TN}^2}{a_{\rm TN}^2}}-1}{\sqrt{1+\frac{4f_{\rm TN}^2}{a_{\rm TN}^2}}+1}.
\end{equation}

Let us concretely give a metric interpolating ${\cal N}_{\rm EH}$ and
${\cal N}_{\rm TN}/{\bf Z}_2$.
As a simplest choice, we use the following metric which is obtained
by simply connecting $(f_{\rm EH},g_{\rm EH})$ and  $(f_{\rm TN},g_{\rm TN})$.
\begin{eqnarray}
&&f(r)=f_{\rm EH}(r),\quad
  g(r)=g_{\rm EH}(r)\quad\mbox{for $0\leq r\leq r_0$}\nonumber\\
&&f(r)=f_{\rm TN}(r-r_1+r_0),\quad
  g(r)=g_{\rm TN}(r-r_1+r_0)\quad\mbox{for $r_0\leq r$},
\end{eqnarray}
where $r_0$ and $r_1$ represent the radial coordinate at the intersection in Figure \ref{gplot.eps} and
are defined by the following junction condition.
\begin{equation}
f_{\rm EH}(r_0)=f_{\rm TN}(r_1),\quad
g_{\rm EH}(r_0)=g_{\rm TN}(r_1).
\label{junction}
\end{equation}
Because our purpose is to show only the existence of a tachyonic mode,
this special choice of the interpolating metric is sufficient.
At $r=r_0$, $f'$ and $g'$ are not continuous
and this manifold is singular.
Indeed the scalar curvature for the metric (\ref{fgmetric})
\begin{equation}
R=-2\frac{(f^2g)''}{f^2g}+\frac{-g^3+4f^2g(1+f'^2)+8f^3f'g'}{2f^4g},
\label{curvature}
\end{equation}
diverges at $r=r_0$.
However, this divergence has a $\delta$-function-like
form and the Lagrangian
\begin{equation}
L=N\int d^4x\sqrt gR,
\label{EHlag}
\end{equation}
is finite.
The overall factor $N$ in (\ref{EHlag})
includes the dilaton factor $e^{-2\phi}$ and the five-dimensional
spatial integral transverse to ${\cal M}_{\SO(3)}$.
Because the interpolating manifold is Ricchi flat for $r\neq r_0$,
and the second term in (\ref{curvature}) is everywhere finite,
only contribution to the Lagrangian comes from the first term
in (\ref{curvature}) at the singularity $r=r_0$
and we obtain
\begin{equation}
V(a_{\rm EH})=-L
 =16\pi^2N[(f^2g)']^{r_0+\epsilon}_{r_0-\epsilon}
 =-16\pi^2N(f(r_0)-g(r_0))^2.
\label{potential}
\end{equation}
In the last step in (\ref{potential}), we used the relations (\ref{killing}).
This is clearly non-positive.
For large $a_{\rm EH}$, $f$ and $g$ at the junction point
approach to $a_{\rm EH}$ and $a_{\rm TN}$, respectively (See Figure \ref{gplot.eps}).
This implies that the potential is unbounded below.
A solution of the junction condition (\ref{junction}) for small $a_{\rm EH}$
is
\begin{equation}
f(r_0)\sim g(r_0)\sim\left(\frac{a_{\rm TN}^2a_{\rm EH}^4}{2}\right)^{1/6},\quad
f(r_0)-g(r_0)\sim\frac{a_{\rm EH}^2}{\sqrt2 a_{\rm TN}}.
\end{equation}
Substituting this into the potential (\ref{potential}),
we obtain the following potential for small $a_{\rm EH}$.
\begin{equation}
V(a_{\rm EH})\sim-8\pi^2N\frac{a_{\rm EH}^4}{a_{\rm TN}^2}.
\label{VrEH}
\end{equation}
In order to determine the mass of this mode,
we need to know the kinetic term of the parameter $a_{\rm EH}$.
Up to a numerical constant, it is determined to be
$L_{\rm kin}\propto N(\partial_t a_{\rm EH}^2)^2$
by a dimensional analysis.
Therefore, the potential (\ref{VrEH}) implies that
the unstable mode parameterized by $a_{\rm EH}$
has a tachyonic mass
\begin{equation}
M^2\propto-\frac{1}{a_{\rm TN}^2}.
\label{M2MSO3}
\end{equation}
Although we cannot determine the numerical coefficient in (\ref{M2MSO3})
from the potential (\ref{VrEH}) because of the artificial choice of the metric,
we can conclude that there exists at least one unstable mode with the non-vanishing
tachyonic mass.
Once we know this, the relation (\ref{M2MSO3}) is immediately obtained
by a dimensional analysis and it is expected that (\ref{M2MSO3}) still
holds if we carry out more detailed analysis of decay modes.

\section{Closed string tachyon condensation}
In the previous section,
we showed in the T-dual picture that a type 0 NS5-brane is unstable
and decays to a bubble of nothing.
It is natural to ask what is responsible for this instability
on the type 0 side.

It is already shown in \cite{ima2} that
there are localized tachyonic modes
on coincident unwrapped NS5-branes in type 0 theory
compactified on ${\bf S}^1$.
This result is obtained by solving the Klein-Gordon equation
for the closed string tachyon field on the background.
In \cite{ima2}, only the small compactification limit is considered
and it is shown that the spectrum of the localized tachyonic modes
is identical to the spectrum of twisted string modes at an
${\bf A}_n$ type singularity, which is T-dual to the NS5-branes.
In the case of single NS5-brane,
there is no tachyonic modes and the lightest localized modes
are massless in the small radius limit\cite{ima2}.

The compactification radius $R$ in type $0$ theory
is related to the parameter $a_{\rm TN}$ by the T-duality relation\cite{BG,ima1}
\begin{equation}
a_{\rm TN}R=\frac{l_s^2}{2}.
\end{equation}
Let us generalize the analysis of tachyonic modes on NS5-branes
given in \cite{ima2} to the case of $R\neq0$.
The metric of NS5-brane solution is
\begin{equation}
ds^2=\sum_{i,j=0}^5\eta_{ij}dx^idx^j+H(dy^2+dr^2+r^2d\Omega_2^2),\quad
e^{2\phi}=\mbox{const}\times H,
\label{NS5sol}
\end{equation}
where $y$ is compactified on ${\bf S}^1$ as $y\sim y+2\pi$.
Because we assume that the compactification radius is very small,
we can use a smeared solution.
For coincident NS5-branes with the total charge $Q$,
the harmonic function $H$ is given by
\begin{equation}
H=R^2+\frac{l_s^2Q}{2r}.
\end{equation}
The free part of the action of the closed string tachyon field is
\begin{equation}
S_{\rm tachyon}=-\int d^{10}x\frac{\sqrt{-g}}{2e^{2\phi}}
\left[(\partial_\mu T)^2+M_T^2T^2\right].
\end{equation}
Let us factorize the
wave function of the tachyon field as
\begin{equation}
T=\psi(r,\theta,\phi)\exp i(k_yy+\sum_{i=0}^5k_ix^i).
\end{equation}
The Klein-Gordon equation for the tachyon field on the NS5-brane solution
(\ref{NS5sol}) is
\begin{equation}
-\Delta_3\psi-(M_6^2-M_T^2)H\psi=-k_y^2\psi,
\end{equation}
where $M_6^2=-\sum_{i,j=0}^5\eta_{ij}k^ik^j$ and
$\Delta_3$ is the Laplacian on ${\bf R}^3$ with the metric
$ds^2=dr^2+r^2d\Omega_2^2$.
Because of the twisted boundary condition, $k_y$ is quantized to be half odd integer.
The Schr\"odinger equation has the same form with that for an electron
in a hydrogen atom and is solved as
\begin{eqnarray}
l_s^2M_6^2&=&-2+4n\left[\sqrt{\frac{k_y^2}{Q^2}+\left(\frac{2nR^2}{Q^2l_s^2}\right)^2}-\frac{2nR^2}{Q^2l_s^2}\right]\nonumber\\
&=&-2+\frac{4n|k_y|}{Q}-\frac{8n^2R^2}{Q^2l_s^2}+{\cal O}\left(\frac{R^4}{l_s^4}\right),
\end{eqnarray}
where we used $M_T^2=-2/l_s^2$.
The multiplicity of states is $n^2$ for each $k\in{\bf Z}+1/2$ and $n=1,2,3,\ldots$.
In the case of single NS5-brane ($Q=1$),
the mass of the lightest states $(k,n)=(\pm 1/2,1)$ is given by
\begin{equation}
M^2
=-\frac{8R^2}{l_s^4}+\frac{1}{l_s^2}{\cal O}\left(\frac{R^4}{l_s^4}\right)
=-\frac{2}{a_{\rm TN}^2}+\frac{1}{a_{\rm TN}^2}{\cal O}\left(\frac{l_s^2}{a_{\rm TN}^2}\right).
\label{closedtachyonmass}
\end{equation}
Thus, there are two localized tachyonic modes in the
NS5-brane background (\ref{NS5sol}).
The tachyonic mass approaches to zero as
$R\propto a_{\rm TN}$ decreases
and is of the same order with (\ref{M2MSO3}).
This strongly suggests that on the type 0 side
what is responsible for the instability of an NS5-brane is
the localized modes of the closed string tachyon field.

\section{Discussions}
In this paper, we analyzed an instability of type 0 NS5-branes
in two ways.
By analyzing the Kaluza-Klein monopole like geometry,
which is T-dual to a type 0 NS5-brane,
we obtained a tachyonic mass of a decay mode proportional
to inverse square of the compactification radius $a_{\rm TN}$.
This coincides up to a numerical coefficient
with the result of the analysis of the closed string tachyon field
on the NS5-brane classical solution.

In our analysis, to simplify the calculation, we used the metric
determined by hand.
Although it was sufficient to show the existence of at least one tachyonic mode,
we need more careful treatment of fluctuation modes to
determine the number of tachyonic modes and the numerical proportional constant in (\ref{M2MSO3}),
and to compare it to the tachyonic mass (\ref{closedtachyonmass})
of the closed string modes on the NS5-brane background.
For this purpose, technique developed
for instanton calculus\cite{constrained,ssbbyinst,susybreaking} may
be useful.

The geometry of ${\cal M}_{\SO(3)}$
is quite similar to a monopole-anti monopole geometry
studied in \cite{decay,nucl} and used for describing an instability of fluxbranes in \cite{Melvin,fluxbrane}.
In \cite{decay,nucl} it is shown that
four-dimensional Euclidean Schwarzschild black hole,
which describes semi-classical instability of a Kaluza-Klein vacuum, is
equivalent to a monopole-anti monopole pair in the Kaluza-Klein theory.
This can be shown as follows.
The isometry of four-dimensional Euclidean Schwarzschild black hole is
$\SO(3)\times\U(1)$, which is identical to that of ${\cal M}_{\SO(3)}$.
If we define ${\bf S}^1$ fiber as orbits of the $\U(1)$ factor of the isometry,
the base manifold is ${\bf R}^3$ with a solid ${\bf S}^2$ omitted.
This is completely the same situation with ${\cal M}_{\SO(3)}$.
On the other hand, if we use a diagonal $\U(1)$ of the $\U(1)$ factor
of the isometry and a $\U(1)$ subgroup of the $\SO(3)$ factor,
we obtain ${\bf S}^1$-bundle over ${\bf R}^3$ with two NUT singularities.
The NUT charges for two singularities have opposite signs.
Therefore, we can regard the geometry as a monopole-anti monopole pair.
When we show that the charges have opposite signs, it is crucial that
the black hole geometry has vanishing first Chern class when it is
represented as an ${\bf R}^2$-bundle over ${\bf S}^2$.
For ${\cal M}_{\SO(3)}$, it is not the case.
In order to represent ${\cal M}_{\SO(3)}$ as a system of
Kaluza-Klein monopoles, we use a $\U(1)$ subgroup of
$\SO(3)$. There are two fixed points under the $\U(1)$
and we obtain ${\bf S}^1$ fibration over ${\bf R}^3$
with two NUT singularities again.
However, at this time, the NUT charges of two singularities are
both $+1$.
Therefore, the geometry is identified with a system of two monopoles.
The expansion of the bubble may be understood as a repulsion
of these two monopoles.

\section*{Acknowledgements}
I would like to thank H.~Takayanagi and T.~Takayanagi
for helpful discussions.
This research is supported in part by
Grant-in-Aid for the Encouragement of Young Scientists
(\#15740140) from the Japan Ministry of Education, Culture, Sports,
Science and Technology,
and by Rikkyo University Special Fund for Research.


\end{document}